\documentclass{article}    
\title{\bf From time inversion to nonlinear QED\footnote{Published on
    {\it Foundations of Physics, Vol. 30, No. 11, 2000}}}
\author{Wei Min Jin\\
{\small Department of
    Physics and Astronomy, State University of New York at Buffalo,}\\
{\small Buffalo, NY 14260-1500, U.S.A.}\footnote{The author has left
  University of Buffalo after graduation.}\\
{\small Electronic mail: wjin1@excite.com}}     
\date{}      
\begin{document}             
\maketitle                   
\begin{center}
\begin{minipage}{135mm}
\baselineskip15pt
\begin{center} {} \end{center}
{In Minkowski flat space-time, it is perceived that time inversion is
unitary rather than antiunitary, with energy being a time vector changing
sign under time inversion. 
The Dirac equation, in the case of electromagnetic interaction, is not
invariant under unitary time inversion, giving rise to a ``Klein paradox''. To   
render unitary time inversion invariance, a nonlinear 
wave equation is  
constructed, in which the ``Klein
paradox'' disappears. In the case of Coulomb interaction, the revised nonlinear equation can be linearized to give energy solutions for 
Hydrogen-like ions without singularity when nuclear number
$Z>137$, showing a reversed energy order pending for experimental tests such as Zeeman effects. In non-relativistic limit, this nonlinear 
equation 
reduces to nonlinear schr\"odinger equation with soliton-like solutions. Moreover, particle
conjugation and 
electron-proton scattering with a nonsingular 
current-potential interaction are discussed. Finally the explicit form of gauge 
function is found, the uniqueness of Lorentz gauge is proven and the
Lagrangian density of quantum electrodynamics (QED) is 
revised as well. The implementation of unitary time 
inversion leads to the ultimate derivation of nonlinear QED.}  
\end{minipage}
\end{center}
\vskip1in
\baselineskip15pt
\eject     

\section* {1. INTRODUCTION}

Decades ago, Wigner [1] first pointed out the significance of time
inversion and later made detailed discussions on antiunitary time inversion
(Wigner [2]). His theory about time inversion is based on a classical motion
picture (Wigner [3]): ``Time inversion \ldots\ replaces every velocity by the opposite
velocity, so that the position of particles at $+$t becomes the same as it
was, without time inversion at $-$t. \ldots'' 

Also well known is Einstein's relativity theory (Einstein [4]) that
completely changes our 
perception of space and time, as best manifested by Minkowski's 
``world-postulate'' (see Minkowski's paper in [4]). Minkowski unifies space and time by introducing time as 
an independent coordinate in addition to three space
coordinates. Usually ``Minkowski space-time'' refers to 
four-dimensional flat space-time in special relativity, in which relativistic
quantum mechanics and quantum electrodynamics (QED) are
established. 

On the development of quantum mechanics and relativity, there always
exist heated debates on the physical meaning of the theories. Dirac however
was indifferent in such debates. Instead he had been
looking for mathematical possibilities to 
reconcile quantum mechanics and relativity. As a consequence, he
founded relativistic quantum mechanics with a relativistic
wave 
equation called ``Dirac equation'' (Dirac [5]). However, the Dirac equation was
also not perfect. In the following year, Klein [6] found a paradox in
the Dirac equation. Now seven decades has gone by, the ``Klein 
paradox'' has still not been resolved mathematically, though it can be
explained away by physical reasoning.

This paper is organized as follows. Given that a correct concept in Newtonian
classical mechanics may not be necessarily correct in Einsteinian
relativistic mechanics, first I am going to show that the above
mentioned classical motion picture is just one of those concepts, correct
classically but not relativistically. Then I will proceed to clarify  
what is supposed to mean to make a time inversion in special relativity.
Based on the general principles of quantum mechanics and special relativity, 
it can be shown that time inversion turns out to be  
a unitary transformation with energy being a time vector changing sign under
time inversion in Minkowski space-time.

With unitary time inversion understood, the next step is to study the
transformation of the Dirac equation. For clarity,
let us name the Dirac equation in the case of electromagnetic interaction as
Dirac EM-equation to distinguish it from the Dirac equation in other
cases. What  
happens is: the Dirac EM-equation is not invariant under unitary
time inversion. As a direct consequence of this non-invariance, the Dirac
EM-equation has a non-symmetric positive-negative energy spectrum with a
mathematical singularity that causes a ``Klein paradox''. Here in this paper,
I do not intend to argue too much about whether or not the ``Klein paradox''
has physical meaning. Rather I hold such a point of view that we would
be better off from the outset without mathematical singularity.

To implement unitary time inversion, I come
up with a revised equation that looks similar to the Dirac EM-equation on
the one
hand, while differs from the Dirac EM-equation in several aspects on the
other. The main difference is: the revised
equation preserves the invariance under unitary time inversion in addition to
the invariance under the other Lorentz transformations.
Consequently, it gives a positive-negative symmetric energy spectrum
without singular crossing point. In the case of Coulomb
interaction, the energy bound states for 
Hydrogen-like ions are solved without singularity when nuclear number $Z>137$, and the order of some energy levels found
opposite to the conventional one, which entails further high
precision tests from experiments like Zeeman effects. 

Another difference is: the Dirac EM-equation, involving the external electromagnetic potential with minimal coupling, is linear on fermion field; while the revised wave equation, involving the
electro-dynamical interaction  
potential, is nonlinear on fermion field. In 
non-relativistic limit, the Dirac EM-equation reduces to the linear
Schr\"odinger equation, while the revised nonlinear wave equation
reduces to the nonlinear Schr\"odinger
equation with soliton-like solutions well known in nonlinear physics. 

Typically, when
the finite size of fermion is considered, the interaction potential can be generalized by a convolution
between four-current and four-potential with an integral over the finite 
four-dimensional size of the fermion. Using such a nonsingular
convolution can avoid the singularity and infinity troubles
in calculating the self-energy of electron and the divergent integrals of
Feynman Diagrams. 

The gauge invariance is to be discussed at the end. To preserve Lorentz
invariance including the invariance under unitary time inversion, it can be shown
that gauge function is of particular form, not arbitrary, and the  
only gauge condition is Lorentz gauge. It is also straightforward
to write a nonlinear 
Lagrangian that derives the Maxwell electromagnetic field equation and the
revised nonlinear fermion field 
equation. With the basic   
principles of quantum field theory, we may then establish a nonlinear QED
that would show many interesting applications in experiment.  What I am
actually up to, is to make necessary 
improvements on QED in its own framework, namely in the language of
space-time, without further drastic 
changes as in superstring theory. So much has been said, let me now turn into more
detailed discussions.

\section* {2. UNITARY TIME INVERSION}
Usually under a coordinate transformation $x'=ax$,
the transformation of wave function in quantum mechanics is defined by
$${\rm A}(a)\Psi(x)=\Psi(ax).\eqno(2.1)$$
But antiunitary time inversion is defined by (see Bjorken
and Drell [7]) 
$${\rm T}_*\Psi(t,{\bf x})=\Psi^*(-t,{\bf x}).\eqno(2.2)$$
where the extra complex conjugation $*$ on wave function is used
to comply with the assumption that under time inversion $(t,{\bf
x})\rightarrow(-t,{\bf x})$, energy and momentum transform as 
$(E,{\bf p})\rightarrow(E,{\bf -p})$.
In this case, the phase of a plane wave 
$\phi={\bf p\cdot x}-Et$ (natural units are used in this
paper),
is not invariant under antiunitary time inversion. This also leads to another
result that the 
imaginary unit $i$ has to be supposed to change to $-i$ under antiunitary
time inversion, while constant $i$ has nothing to do with time. 

In quantum mechanics, the group velocity of a wave packet is
expressed by $d{\bf x}/dt=dE({\bf p})/d{\bf p}$.
By making a positive non-relativistic energy assumption
$E({\bf p})={\bf p}^2/2m$, one can then get a classical correspondence 
$d{\bf x}/dt={\bf p}/m={\bf v}$,
from which the classical motion picture comes. The point is: in the
derivation of classical mechanics velocity from quantum mechanics
group velocity, the positive energy assumption has been taken, which may not be correct
under time inversion in special relativity. It would be logically
odd to   
discuss symmetry under time inversion based on a truncated
classical motion picture that has already broken the symmetry.

Now that quantum mechanics
is more general than classical mechanics, it is better not to impose any
properties for time inversion classically until we obtain certain results from
quantum mechanics. As we understand, time 
inversion is nothing but a kind of coordinate transformation in 
space-time, and it is natural to adapt the general definition (2.1) for time
inversion rather than to follow the truncated classical motion
picture. Therefore putting the general principles of quantum mechanics in the
first place, we define time inversion as: 
$${\rm T}\Psi(t,{\bf x})=\Psi(-t,{\bf x}).\eqno(2.3)$$

For a plane wave $\Psi(t,{\bf x})=C({\bf p},E)\exp[i({\bf p\cdot
x}-Et)]$, 
it is clear that the expectation value of space momentum operator 
${\bf\hbox{\^p}}=-i\partial_{\bf x}$ remains the same and that of energy
operator  
${\hbox{\^E}}=i\partial_t$ changes sign under time inversion. Any wave
function can be 
Fourier-transformed into a combination of plane waves. 
So it is easy to check these results hold for any kind of matter
waves.

In 4-d space-time, the space-time interval squared 
$\Delta X_\mu\Delta X^\mu=g^{\mu\nu}\Delta X_\mu\Delta
X_\nu$ is Lorentz invariant, where 
$g^{00}=-g^{ii}=1(i=1,2,3)$ and $g^{\mu\nu}=0(\mu\not=\nu)$.
Mathematically if contravariant four-coordinate transforms as
$X'^\mu=a^\mu_\nu X^\nu$, 
then covariant four-coordinate transforms in the way $X'_\mu=a_\mu^\nu X_\nu$.
Since $\Delta{X'_\mu}\Delta{X'^\mu}=\Delta{X_\mu}\Delta{X^\mu}$,
we get $a^{-1}=ga^\sim g$ where $a^\sim$ is the transpose of $a$, leading to 
$\det(a^{-1})=\det(a)=\pm 1$
for proper and improper Lorentz transformations respectively. 
On the other hand, the phase of a plane wave $\phi=-P_\mu X^\mu$
characterizing the physical correlation of space-time world-points, is also
invariant under homogeneous Lorentz transformations, therefore four-momentum must transform in a covariant way $P'_\mu=a_\mu^\nu P_\nu$.
In the case of time inversion $(t,{\bf
x})\rightarrow(-t,{\bf x})$, it is naturally found an energy-momentum
transformation 
$$(E,{\bf p})\rightarrow(-E,{\bf p}).\eqno(2.4)$$
 
Moreover, if a plane wave with a constant four-momentum $P_\mu=(E,{\bf p})$
travels from space-time point 1 to 2, the phase difference between them 
$\Delta\phi=\phi_2-\phi_1$
describing the causality of this process, is an invariant
under all inhomogeneous Lorentz transformations $X'^\mu=a^\mu_\nu X^\nu+b^\mu$.
The phase differences between points 1 and 2 before and after space-time
inversions can be written down as: 
$\Delta\phi={\bf p\cdot(x_2-x_1)}-E(t_2-t_1)$ and
$\Delta\phi'={\bf p'\cdot(x'_2-x'_1)}-E'(t'_2-t'_1)$.
Since $\Delta\phi'=\Delta\phi$, we
obtain $\bf p'=-p$ under space inversion $\bf
x'_2-x'_1=-(x_2-x_1)$, and $E'=-E$ under time inversion
$t'_2-t'_1=-(t_2-t_1)$. It is such a Lorentz invariant physical phase space that
determines the unique energy-momentum transformation $(E,{\bf
p})\rightarrow(-E,{\bf p})$ under time inversion. 

Actually four-vector momentum $P_\mu=(E,{\bf p})$ consists
of time component $E$ and space component $\bf p$. While momentum $\bf p$ is
usually known as a space vector changing sign under space inversion, 
energy $E$ may be similarly regarded as a time vector changing sign under time
inversion. The idea of uncertainty principles in quantum
mechanics: $\Delta{\bf p}\Delta{\bf x}\sim 1$ and $\Delta E\Delta t\sim 1$,
is that the strict measurement of momentum and position as well as that of
energy and time can not be taken simultaneously since they do not commute,
but they do have close relationship with each other. The determination of
momentum is tightly related to that of position not to that of
time. Similarly the measurement of energy is directly connected to that of
time not to that of position. This one-to-one correspondence reveals the
following fact that 4-d space-time and energy-momentum, combined
into a Lorentz invariant phase space, are the reciprocal systems
relatively to each other 
$$P_\mu=i\partial_\mu,\eqno(2.5)$$
with a commutator relation $[X^\mu, P_\mu]=-i$.
 
The Fourier transformation used in quantum mechanics also shows a good
example of this one-to-one correspondence. According to this
correspondence, we prefer to call 4-d energy-momentum system, the
``reciprocal world'' of space-time. Generally speaking, this is only a
question of using different representations or interchangeable
languages. There is nothing special, in the 
sense of relativistic covariance, that producing an inversion in space-time
world is equivalent to making a corresponding inversion in its energy-momentum
reciprocal world by the hint of (2.5). This idea now enables us to understand
why momentum switches sign under space inversion while energy reverses sign
under time inversion.  

The physical picture becomes clear if in a ``local framework'', we do not
impose absolute future and 
past concepts by setting a standard clock A but rather reverse time axis by
setting another clock B running counterclockwise. If a
plane wave is propagating toward the future with a positive (or negative)
frequency by the standard clock A, then it can also be equivalently looked
upon as propagating toward the past with a negative (or positive) frequency
by the other clock B. On the principle of special relativity, it is arbitrary
in setting clocks in locally flat space-time. Therefore this plane wave may have
either positive or negative energy depending on which clock we use. 

By the
correspondence between space-time and energy-momentum worlds, if we indeed
want to introduce a time inversion concept in Lorentz group, then it is
natural to introduce a corresponding ``energy inversion'' concept to map these
two worlds completely. Once again we would like to emphasize, based on what we
try to clarify here from the intrinsic structures of space-time world and
energy-momentum reciprocal world, and from the general principles of special
relativity and quantum mechanics, that in Minkowski flat space-time, energy
is not a scalar any more, it is a ``time vector'' changing sign under time
inversion.

Similar to space inversion, by definition (2.3) it is easy to
prove: (a) T is linear; (b) ${\rm T}^2=1$; (c) ${\rm T}^*=$T; (d) ${\rm
T}^\sim=$T. So
time inversion operator is unitary 
$${\rm T}^{-1}={\rm T}={\rm T}^\dagger.\eqno(2.6)$$
To find its eigenvalues let T$\Psi=\lambda\Psi$. From relation (b) we get 
$\lambda=\pm 1$
representing even and odd ``time parities'' respectively. 

For a free relativistic particle with an energy-momentum relation $E^2-{\bf
p}^2=m^2$,  
the common eigenfunctions of momentum operator ${\bf\hbox{\^p}}=-i\partial_{\bf x}$ and energy operator
$\hbox{\^E}=i\partial_t$ are expressed by 
$\Phi_\pm(t,{\bf x})=C({\bf p},E)\phi_{\bf p}({\bf x})\exp(\pm iEt)$. Obviously they are
not the eigenfunctions of time inversion operator T 
since T does not commute with \^E: $[{\rm T},\hbox{\^E}]\not=0$.
Now make linear combinations: $\Psi_{+}=(\Phi_{+}+\Phi_{-})/2$ and $\Psi_{-}=
(\Phi_{+}-\Phi_{-})/2i$,
which are the common eigenstates of {\bf\^p} and T but not those of \^E 
any more. Generally any state wave function can be divided by 
$\Psi=\Psi_{+}+\Psi_{-}$,
where the even and odd time parity eigenstates are 
$\Psi_\pm=(1\pm {\rm T})\Psi/2$ respectively.
And any operator can be expressed by $\hbox{\^W}=\hbox{\^W}_{+}+\hbox{\^W}_{-}$,
where $\hbox{\^W}_\pm=(\hbox{\^W}\pm{\rm T}\hbox{\^W}{\rm T})/2$
and T$\hbox{\^W}_\pm{\rm T}=\pm\hbox{\^W}_\pm$.
Here $\hbox{\^W}_{+}$ represents the even time parity operator such as 
momentum $\hbox{\bf\^p}=-i\partial_{\bf x}$ and $\hbox{\^W}_{-}$ represents
the odd time parity operator such as energy $\hbox{\^E}=i\partial_t$.

\section* {3. DIRAC EQUATION REVISITED}
At the end of 19th century, Zeeman [8] detected the splitting of spectral
lines under the influence of external magnetic fields, revealing the
existence of electron spin. To explain the Zeeman effect, the imagination of
electron structure has been made: electron has a magnetic moment and thus has
an intrinsic spin angular momentum instead of a classical point particle
(Lorentz [9]). It has been clarified that electron spin relates to the extra
degrees of freedom and can be well described by a two component complex
variable called ``spinor''. One is then faced with such a typical question:
how to construct a minimal complete set of variables in the combination of
space-time and intrinsic spin. In Dirac's mind (Dirac [10]), spinors, like
tensors, are geometrical objects, yielding covariant transformations with
respect to Minkowski space-time, and specifically under Lorentz
transformations, the Dirac equation is to obey the covariance principle
of special relativity. Along this line of thought, we may say the whole
domain of space-time and intrinsic spin is Lorentz invariant, and name it ``common space'' for clarity. 

A common variable is now defined by
$$\Omega=\gamma^\mu X_\mu,\eqno(3.1)$$
where matrices $\gamma^\mu (\mu=0,1,2,3)$, in the representation of Bjorken and Drell [7],  are ascribed to the spin  
geometrical factors. The common
variable constructed in this way is a function of space-time coordinates and
also a metrical quantity. The common variable interval squared
becomes   
$\Delta\Omega^2=\Delta X_\mu\Delta X^\mu$, leading to the anticommutation relations of Dirac matrices
$\Bigl\{\gamma^\mu,\gamma^\nu\Bigr\}=2g^{\mu\nu}$. Moreover, the derivative
with respect to common variable can be deduced as 
$$\partial_\omega=({1\over{\partial_\mu\Omega}})\partial_\mu=\gamma^\mu
\partial_\mu,\eqno(3.2)$$
and a momentum operator in common space is naturally introduced as 
$$P_\omega=i\partial_\omega=\gamma^\mu P_\mu.\eqno(3.3)$$
By (3.1) and (3.3) it is straightforward to prove a commutator relation 
$$[\Omega, P_\omega]=-4i.\eqno(3.4)$$
If noticing that the common momentum squared of a free particle with mass $m$
is constant $P^2_\omega=E^2-{\bf p^2}=m^2$, 
we find $P_\omega$ has two eigenvalues $\pm m$. If choosing $+m$
eigenvalue, we arrive at the Dirac equation of free spin one-half particle 
$$\gamma^\mu P_\mu\Psi(x)=m\Psi(x),\eqno(3.5)$$

The Dirac equation, on the principle of special relativity, is required
to be invariant under Lorentz transformation ${\rm L}_\omega$ in common space,
which is a direct product of spinor one ${\rm L}_s$ and coordinate one ${\rm
L}_c$: ${\rm L}_\omega={\rm L}_s{\rm L}_c$. Making ${\rm L}_\omega$ on both
sides of (3.5) and noting that ${\rm L}_s$ commutes with
space-time vectors and ${\rm L}_c$ commutes with spinor vectors, we have
$${\rm L}_s\gamma^\mu {\rm L}_s^{-1}{\rm L}_cP_\mu
{\rm L}_c^{-1}{\rm L}_\omega\Psi(x)=m{\rm L}_\omega\Psi(x).\eqno(3.6)$$ 
Since momentum $P_\mu$ is a covariant four-vector in space-time,
$${\rm L}_cP_\mu {\rm L}_c^{-1}=a_\mu^\nu P_\nu,\eqno(3.7)$$ 
matrices $\gamma^\mu$ must be correspondingly contravariant in spinor space, 
$${\rm L}_s\gamma^\mu {\rm L}_s^{-1}=a^\mu_\nu\gamma^\nu.\eqno(3.8)$$

The above discussions also hold for two special cases.
First, there is a unitary space inversion in common space ${\rm
S}_\omega={\rm S}_s{\rm S}_c$,
where ${\rm S}_s=\gamma^0$ is a unitary space inversion in spinor space: 
$${\rm S}_s\gamma^0{\rm S}_s^{-1}=\gamma^0,\;
{\rm S}_s\mbox{\boldmath$\gamma$}{\rm S}_s^{-1}=-\mbox{\boldmath$\gamma$},\eqno(3.9)$$
and ${\rm S}_c$ is a unitary space inversion in space-time, defined as usual in
quantum mechanics. Second, there is a unitary time inversion in common space 
${\rm T}_\omega={\rm T}_s{\rm T}_c$,
where ${\rm T}_s=\gamma^1\gamma^2\gamma^3$ is a unitary time inversion in
spinor space:  
$${\rm T}_s\gamma^0{\rm T}_s^{-1}=-\gamma^0,\;
{\rm T}_s\mbox{\boldmath$\gamma$}{\rm
T}_s^{-1}=\mbox{\boldmath$\gamma$},\eqno(3.10)$$ 
and ${\rm T}_c$ is a unitary time inversion in space-time, defined by (2.3). 

When a spin one-half particle is put into an external Maxwell electromagnetic
field, the Dirac EM-equation with minimal coupling can be written down in a
general form (Dirac [5] [11]): 
$$\gamma^\mu(P_\mu-eA_\mu)\Psi(x)=m\Psi(x),\eqno(3.11)$$
where $A_\mu=(\Phi,{\bf A})$ 
is the space-time four-potential of electromagnetic field, proposed by Minkowski
[12]. Under time inversion, time component $\Phi$ does not change, while
space vector $\bf A$ is supposed to change sign. The Minkowski four-potential
constructed in this way is therefore ``time-anticovariant'' under time
inversion:
$$(\Phi,{\bf A})\rightarrow(\Phi,-{\bf A}).\eqno(3.12)$$ 

If four-momentum $P_\mu$ is also ``time-anticovariant'', one may make
spinor space yield a sort of transformation to keep the equation invariant as
done in conventional theory. However from what
we have shown in the last section, four-momentum 
is not ``time-anticovariant'', but rather ``time-covariant'' (2.4), just as
it is 
covariant under all the other Lorentz transformations. Here exists a
nontrivial dilemma that the summation of ``time-covariant'' four-momentum 
and ``time-anticovariant'' Minkowski four-potential multiplied by
charge and a negative sign is neither
time-covariant nor time-anticovariant, keeping in no way the Dirac EM-equation
invariant under time inversion.  

This dilemma gives rise to an unsymmetrical positive-negative
energy spectrum when we solve this equation. Consequently when the electric 
potential $\Phi$ such as Coulomb potential is strong enough, the positive and
negative energy spectra 
will be shifted to overlap, leading to a mathematical singularity difficulty
- ``Klein paradox'' (Klein [6]). Particularly, for a Hydrogen-like ion there
is no 
real bound energy solution but appears a singularity when the nuclear number
of the ion is larger than 137. 

At first, Dirac could hardly understand why his equation has negative
energy solutions that seem ``nonphysical''. To find an explanation,
Dirac constructed a hole theory with a vacuum of fully filled negative energy
states (Dirac [13]). The predicted positron was
found a few years later (Anderson [14]). Despite Dirac's profound prediction of
antiparticles, the hole theory has   
difficulties such as infinite density, infinite negative energy
and vacuum fluctuation in the vacuum state. Based on the hole theory, the 
``Klein paradox'' is explained in terms of 
vacuum polarization accompanying particle-antiparticle pair production
and annihilation under strong fields.  
Along this line of work, one has been trying to explain
away the ``Klein paradox'' rather than to avoid it. Eventually, one is
led to such a conclusion: ``\,`Klein paradox' is not a paradox''. 
      
Just as in classical mechanics, energy in the Dirac hole theory is
still viewed as a 
scalar: namely negative energy is always ``lower'' than positive energy.
If energy in special relativity is a frame-dependent
``time vector''  
rather than just a scalar, as we have shown earlier, then it makes little
sense to insert 
negative energy into vacuum or explain the ``Klein paradox'' as vacuum
polarization. There should exist a more fundamental solution for these
problems. In the next section, we would like to provide a
down-to-earth approach, to remove the ``Klein paradox'' by revising
the Dirac EM-equation. It sounds a little radical. Indeed this can not be done without the
understanding that time inversion is a unitary transformation and energy is a
``time vector'' changing sign under time inversion in Minkowski space-time.
 
\section* {4. NONLINEAR EQUATION RECONSTRUCTED}
As shown in \S 2, energy is a time vector changing sign under time
inversion. Therefore the whole energy spectrum we obtain from whatever
relativistic wave equation we might construct, must be positive-negative
symmetric. This is, on the other hand, a necessary check if the constructed
equation is truly invariant under unitary time inversion or not. By looking
into the Dirac EM-equation, we find that the non-symmetric energy spectrum is
caused by a matrix $\beta$ in front of electric potential $\Phi$. If $\beta$
is removed there, the equation becomes invariant under unitary time
inversion. But it is no longer invariant under the other proper Lorentz
transformations, if $\Phi$ and {\bf A} are still external potentials. After
careful checking, we realize that interaction potentials have to be utilized
in lieu of external potentials in a special way, to render the invariance under all Lorentz transformations. The following equation is what we get:      
$$(\gamma^\mu P_\mu-e\Phi^{\rm I}+e\mbox{\boldmath$\gamma$}\cdot
{\bf A^I})\Psi(x)=m\Psi(x),\eqno(4.1)$$
where $\Phi^{\rm I}$ and $\bf A^I$ are the interaction potentials
related to the external potentials $\Phi$ and $\bf A$. We are going to derive
their relations by considering Lorentz transformations, among which time
inversion is defined in the unitary way we presented earlier. 

As in \S 3,
one can see this modified equation is invariant under unitary space inversion
${\rm S}_\omega=\gamma^0{\rm S}_c$ and time inversion
${\rm T}_\omega=\gamma^1\gamma^2\gamma^3{\rm T}_c$ where ${\rm T}_c$ is defined by (2.3), by
assuming that scalar potential $\Phi^{\rm I}$ does not change sign under
space-time 
inversions while space vector potential $\bf A^I$ changes sign under space
inversion but not under time inversion. These assumptions can be directly
verified after we derive the explicit forms of $\Phi^{\rm I}$ and $\bf A^I$
in terms of $\Phi$ and $\bf A$. 

Below we are going to prove that the modified equation (4.1) does  
give a symmetric positive-negative energy spectrum, to enhance our confidence
that this equation is indeed invariant under unitary time inversion.
From the equation (4.1), the energy operator is expressed in the way
$$\hbox{\^E}=\rho_1\mbox{\boldmath$\sigma$}\cdot({\bf
p-eA^I})+\rho_3(m+e\Phi^{\rm I}),\eqno(4.2)$$ where 
$$\rho_1=\left(\begin{array}{cc}0&I\\
I&0\end{array}\right),\rho_2=\left(\begin{array}{cc}0&-iI\\
iI&0\end{array}\right),\rho_3=\left(\begin{array}{cc}I&0\\
0&-I\end{array}\right),\eqno(4.3)$$
form a set of $4\times4$ matrices analogous to $2\times2$ Pauli matrices
$\sigma_i(i=1,2,3)$. Here $I$ is a $2\times2$ unit matrix. To diagonalize
(4.2), take a look at the square of $\hbox{\^E}$
$$\hbox{\^E}^2=({\bf
p-eA^I})^2-e\mbox{\boldmath$\sigma$}\cdot(\mbox{\boldmath$\nabla$}\times{\bf
A^I}+{\bf
A^I}\times\mbox{\boldmath$\nabla$})+(m+e\Phi^{\rm
I})^2-e\rho_2\mbox{\boldmath
$\sigma$}\cdot[\mbox{\boldmath$\nabla$},\Phi^{\rm
I}].\eqno(4.4)$$ 
Making a  unitary transformation (${\rm U}^{-1}={\rm U}^\dagger$) 
$${\rm U}={1\over{\sqrt{2}}}\left(\begin{array}{cc}I&iI\\
iI&I\end{array}\right),\eqno(4.5)$$
on both sides of (4.4) we can diagonalize it by using 
$${\rm U}\rho_2{\rm U}^{-1}=-\rho_3,\eqno(4.6)$$
to the following form 
$$\hbox{\^E}'^2={\rm U}\hbox{\^E}^2{\rm U}^{-1}=\left(\begin{array}{cc}\hbox{\^E}_u^2&0\\
0&\hbox{\^E}_l^2\end{array}\right)$$
$$=({\bf
p-eA^I})^2-e\mbox{\boldmath$\sigma$}\cdot(\mbox{\boldmath$\nabla$}\times{\bf
A^I}+{\bf A^I}\times\mbox{\boldmath$\nabla$})+(m+e\Phi^{\rm
I})^2+e\rho_3\mbox{\boldmath$\sigma$}\cdot[\mbox{\boldmath$\nabla$},\Phi^{\rm
I}].\eqno(4.7)$$ 
If supposing the following solutions 
$$\hbox{\^E}_u^2\chi_u=E_u^2\chi_u,\;\hbox{\^E}_l^2\chi_l=E_l^2\chi_l,\eqno(4.8)$$
we can rewrite (4.7) as 
$$\hbox{\^E}'^2=\sum_\chi(|\chi_u>E_u^2<\chi_u|+|\chi_l>E_l^2<\chi_l|).\eqno(4.9)$$
Transforming back to 
$$\psi_u={\rm U}^{-1}\chi_u,\;\psi_l={\rm U}^{-1}\chi_l,\eqno(4.10)$$
we get 
$$\hbox{\^E}^2=\sum_\psi(|\psi_u>E_u^2<\psi_u|+|\psi_l>E_l^2<\psi_l|).\eqno(4.11)$$
It is straightforward to check that the energy operator turns out to be 
$$\hbox{\^E}=\sum_{\psi,\lambda}(|\psi_u>\lambda E_u<\psi_u|+|\psi_l>\lambda
E_l<\psi_l|),\eqno(4.12)$$
with $\lambda=\pm 1$, which indeed gives a symmetric positive-negative energy
spectrum. 

To keep (4.1) invariant under the proper Lorentz transformations  
$$\Psi'={\rm L}_p\Psi,\overline\Psi'=\overline\Psi {\rm L}_p^{-1},\eqno(4.13)$$
where $\overline\Psi=\Psi^\dagger\gamma^0$, we need to let the following term:  
$$\overline\Psi(\Phi^{\rm I}-\mbox{\boldmath$\gamma$}\cdot{\bf
A^I})\Psi\equiv\overline\Psi\Psi J^\mu A'_\mu, \eqno(4.14)$$
be invariant, where 
$$J_\mu=\overline\Psi\gamma_\mu\Psi,\eqno(4.15)$$ 
and 
$$A'_\mu=({\Phi^{\rm I}\over{\Psi^\dagger\Psi}},{{\bf
A^I}\over{\overline\Psi\Psi}}).\eqno(4.16)$$
Since $\overline\Psi\Psi$ is a proper Lorentz
scalar and $J_\mu$ is a proper Lorentz four-current, $A'_\mu$ must
be a proper Lorentz four-vector as well. If we define the following
correspondence between the interaction and external potentials: 
$$\Phi^{\rm I}=(\Psi^\dagger\Psi)\Phi,\eqno(4.17a)$$
$${\bf A^I}=(\overline\Psi\Psi){\bf A},\eqno(4.17b)$$
then we see $A'_\mu$ becomes the Minkowski four-potential
$A_\mu=(\Phi,{\bf A})$. 

Considering a four-current $eJ_\mu$ as a source of four-potential
$A_\mu$, we have a Poisson equation 
$$\partial_\rho\partial^\rho A_\mu=eJ_\mu,\eqno(4.18)$$
which is the Maxwell equation under a Lorentz gauge 
$$\partial^\mu A_\mu=0,\eqno(4.19)$$
with a current continuity equation derivable from (4.1)
$$\partial^\mu J_\mu=0.\eqno(4.20)$$
This whole set of the Maxwell equations is invariant under the
proper Lorentz transformations as well as unitary space-time
inversions, although the Minkowski four-potential and four-current 
are ``both'' time-anticovariant.  

Take a look at an example: if an electron feels the external force driven by
a proton, from the correspondence (4.17) and the Maxwell equation (4.18) we
get the following potential expressions 
$$\Phi^{\rm I}(x)=\Psi_e^\dagger(x)\Psi_e(x)\int
d^4x'[e_p\Psi_p^\dagger(x')\Psi_p(x')]G(x,x'),\eqno(4.21a)$$
$${\bf A^I}(x)=\overline\Psi_e(x)\Psi_e(x)\int
d^4x'[e_p\overline\Psi_p(x')\hbox{\boldmath$\gamma$}\Psi_p(x')]G(x,x'),\eqno(4.21b)$$
where Green's function $G(x,x')$ satisfies 
$$\partial_\rho\partial^\rho G(x,x')=\delta^{(4)}(x-x').\eqno(4.22)$$
Hence $\Phi^{\rm I}$ and $\bf A^I$, which depend on not only the wave function of
driving source $\Psi_p(x')$ but also that of testing body $\Psi_e(x)$, are
the interaction potentials between the driving source and testing
body, gaining different meaning from the background potentials $\Phi$ and
$\bf A$ driven only by external sources. Though the Minkowski
four-potential has 
been successfully used in the Maxwell equation (4.18), the situation in the
wave equation (4.1) is different: the interaction potentials $\Phi^{\rm I}$ 
and $\bf A^I$ between the spin one-half particles and the external
electromagnetic 
fields, should be taken into account, even if $\Phi^{\rm I}$ and ${\bf A^I}$ do
not yet form a covariant four-vector. While a time-covariant interaction
four-potential will be derived in \S 8. Particularly in the linear
approximation, the
equation (4.21a) gives a point-like static Coulomb potential $e_e\Phi^{\rm
I}=-\alpha/r$ as we expect. 

From the relations (4.14) through (4.17), the equation (4.1) can also be
alternatively written down as 
$$\gamma^\mu P_\mu\Psi(x)=(m+eJ^\mu A_\mu)\Psi(x).\eqno(4.23)$$
Here $eJ^\mu A_\mu$ happens to be the well-known
electro-dynamical interaction potential. So unlike the Dirac EM-equation, the
constructed  
equation is a nonlinear wave equation, which, in most cases, does  
not have exact explicit analytical solutions. This result is actually
a consequence of the Maxwell equation, in which four-potential
transforms in the same way as four-current under the Lorentz transformations, leading to the appearance of a Lorentz invariant electro-dynamical 
interaction potential in (4.23). 

Originally the Dirac EM-equation was set up to deal with the
interaction between a 
point-like electron and an external electromagnetic field, while the internal
size of electron was ignored. In contrast, the nonlinear equation (4.23) may become valid for extended electron, when the interaction  
potential is expressed by a convolution between four-current and four-potential      
$$V(x)=eJ^\mu\ast A_\mu=e\int d^4x'J^\mu(x')A_\mu(x-x'),\eqno(4.24)$$
with an integral over the finite 4-d size of extended
electron. Using such a nonsingular convolution 
can avoid the infinity troubles like the infinite self-energy
of electron and divergent integrals in Feynman diagrams, caused by the 
interaction of singular point particles. 

If a fermion also participates in other types of interactions, more Lorentz
invariant interaction terms should be added into $V(x)$. We may just as well
generalize (4.23) by the following form: 
$$i\gamma^\mu\partial_\mu\Psi(x)=[m+V(x)]\Psi(x),\eqno(4.25)$$
where $V(x)$ representing the sum of interaction terms must be invariant
under all the Lorentz transformations. This equation is basically a nonlinear
equation since the interaction $V(x)$ depends on fermion fields. To
completely solve the problem, one needs to establish more equations for
intermediate bosons by gauge theory. Of course, this is only true in
principle. When more complicated interactions are 
involved, solving a complete set of nonlinear equations
are very impractical and much unnecessary, and further realizable
techniques are required. 

\section* {5. ENERGY ORDER OF HYDROGEN-LIKE IONS}
Though (4.1) is a nonlinear equation without exact analytical solutions, we
can still make a fair 
approximation that the interaction potentials $\Phi^{\rm I}$ and $\bf A^I$
are slowly-changing functions of particle wavefunctions on account of the
small scale of particle itself compared with the long-range electromagnetic
interaction. Using this approximation and noticing the differences
between upper larger components and lower smaller components of wave
function, we obtain the non-relativistic Hamiltonian of four major terms after
taking $\bf A^I=B^I\times r$/2 which is valid in the atomic range where $\bf B^I$
is approximately constant:
$$H_{NR}={{{\bf p}^2}\over{2m}}+e\Phi^{\rm I}-{e\over{2m}}{\bf
k}\cdot(\mbox{\boldmath$\nabla$}\times{\bf A^I})-{e\over{2m^2r^2}}({\bf
r}\cdot\mbox{\boldmath$\nabla$}\Phi^{\rm I})({\bf s\cdot l}),\eqno(5.1)$$ 
here ${\bf k}={\bf l}+\mbox{\boldmath$\sigma$}$, while $\bf l=r\times p$ is
the orbital 
angular momentum, ${\bf s}=\mbox{\boldmath$\sigma$}/2$ is the spin, and the
total angular momentum is $\bf j=l+s$. In (5.1), the first term is the
kinetic
energy; the second is the electric potential; 
the third shows the magnetic Zeeman effect; the forth represents the spin-orbit
coupling. As we can see, the first three terms are formally the same as the
conventional results although $\Phi^{\rm I}$ and $\bf A^I$ refer to the 
interaction potentials not the pure external potentials. The forth spin-orbit 
coupling term has a different sign from convention, showing a 
reversed order of splitting energy levels. 

With the non-relativistic Hamiltonian (5.1), we can write a wave
equation $i\partial_t\psi=H_{NR}\psi$, which is kind of ``nonlinear
Schr\"odinger equation'' since the interaction potentials $\Phi^{\rm
  I}$ and $\bf A^I$ depend on the wave 
functions. The nonlinear Schr\"odinger equation has certain soliton-like
solutions, and has been applied in many aspects of nonlinear physics. 
However in the derivation of the above non-relativistic
Hamiltonian, the positive-energy assumption has been taken. This
implies: a  
wave equation with a non-relativistic Hamiltonian is not invariant under
unitary time inversion. In other words, the results obtained by applying time
inversion in 
a truncated non-relativistic wave equation like the Schr\"odinger
equation may not be correct. 

As an example, we can solve the nonlinear relativistic wave equation
(4.1) for the single-electron model of 
Hydrogen-like ions under the linear approximation that there is a point-like
Coulomb potential $e\Phi^{\rm I}=-Z\alpha/r$, but no vector magnetic
potential $e{\bf A^I}=0$. In this model, (4.1) reduces to
$$(\gamma^\mu P_\mu+{Z\alpha\over r})\Psi(x)=m\Psi(x),\eqno(5.2)$$
which differs from the Dirac equation by a matrix $\beta$ on the
Coulomb term. A more general form than (5.2) for scalar central
potential ($C/r$) has been investigated in detail (Greiner [15]
[16]). Following the standard procedures of solving this type of
equations, we arrive at the bound state energy solutions of (5.2): 
$$E_{nj}=\pm m\Bigl\{1-{{(Z\alpha)^2}\over{[n-(j+1/2)+
\sqrt{(j+1/2)^2+(Z\alpha)^2}]^2}}\Bigr\}^{1/2},\eqno(5.3)$$
which gives a symmetric positive-negative energy spectrum and does not have any
difficulty when $Z\alpha>1$. 

If $Z\alpha\ll 1$, we can expand the positive $E_{nj}$ in powers of $Z\alpha$
$$E_{nj}=m[1-{{(Z\alpha)^2}\over{2n^2}}+{{(Z\alpha)^4}\over{2n^3}}({1\over{j+1/2}}-{1\over{4n}})]+O[(Z\alpha)^6].\eqno(5.4)$$
Where the first term is the rest mass of electron, the second term is the
classical quantum mechanics result, and the third relativistic term is
different from the conventional one by a crucial sign change. Although the
fine structure splitting energy happens to be the same as usual 
$$\Delta E_n(j_1\rightarrow
j_2)={{m(Z\alpha)^4}\over{2n^3}}|{1\over{j_1+1/2}}-{1\over{j_2+1/2}}|,\eqno(5.5)$$
the energy order for quantum number $j$ with fixed $n$ is changed. Our conclusion
is: the smaller the total angular momentum $j$, the larger the positive
energy $E_{nj}$ in a certain level $n$ of Hydrogen-like ions. For
example, the
energy level $2P_{1/2}$ is higher than the energy level $2P_{3/2}$ in the
doublet $2P_{1/2}-2P_{3/2}$ of Hydrogen-like ions, contrary to the
conventional order. The common practice in experiment is: first detect
a spectrum of energy levels, then arrange these energy levels in the
order derived from the Dirac equation.   
Up till now, the energy order of the doublet of Hydrogen-like ions, has not been precisely verified in experiment as far as we know. Let us examine if it is possible to double check it by experiment.  

The fine structures of the lines $\rm H_\alpha(\lambda6563,n=3\rightarrow2)$
and $\rm H_\beta(\lambda4861,n=4\rightarrow2)$ in the Balmer series of
Hydrogen have been carefully investigated (Lewis and Spedding [17]; Spedding
{\it et al.} [18]), of which no more than two strong 
components can be resolved within experimental accuracy. This type of
experiments, however, can only show the energy difference between two levels,
i.e., $2P_{1/2}-2P_{3/2}$ of Hydrogen atom. It can not, by itself, indicate
the energy order: which line corresponds to which transition. The fine
structure of the line $\lambda4686(n=4\rightarrow3)$ of singly ionized Helium
$\rm He^{+}$ has also been studied in detail (Paschen [19]). However, if the
new energy 
order is assigned to those energy levels, the fit to the 
experimental data is not worse in consideration of
experimental accuracies. On the other hand, the Paschen-Back [20] effect by
applying strong magnetic fields shows a symmetric triplet optical spectrum,
and does not make any difference if the order of energy levels is reversed. 

On the other hand, the Zeeman effects of multi-electron atoms and ions have
been vastly investigated (White [21]; Kuhn [22]), which show both ``normal
order'', the 
bigger $j$ the higher energy level, and ``abnormal order'', the smaller $j$ the
higher energy level. When a number of electrons are involved in
multi-electron atoms or ions, more complicated effects must be taken into
account, for example, electrostatic screening, orbit penetration and
Fermi-Dirac statistics properties of multi-electron systems, and more
complete equations like the self-consistent Hartree-Fock equations
need to be set 
up. They are beyond the scope that the single-electron equation
can describe. 

In principle, only the Zeeman effects for Hydrogen-like ions under weak
magnetic fields can directly and clearly give us the answer of the
energy order of Hydrogen-like ions, since the
asymmetric spectral lines for different transitions may be presented (White
[21]; Kuhn [22]). The principal 
fine structure doublet splitting ($2P_{1/2}-2P_{3/2}$) of Hydrogen atom is
less than half wave number, hence it is very difficult to detect more
splitting levels under weak magnetic fields in such a narrow band with good
resolution. For this reason, the Zeeman effect of Hydrogen atom has not been
observed (White [21]), and the energy 
order of the doublet $2P_{1/2}-2P_{3/2}$ in Hydrogen is still an
unsolved puzzle 
in experiment. If high resolution (say, one tenth of wave number) in
analyzing spectral lines is realized, the detection of the Hydrogen
Zeeman effect will be easier. For the other Hydrogen-like ions such as
$\rm He^{+}, 
Li^{++}, Be^{+++}$, etc., having bigger fine structure splitting energies
($\propto Z^4$) as seen in (5.5), the Zeeman effects under weak magnetic
fields seem detectable, even if most of the simple transitions
($n=2\rightarrow1$, or $3\rightarrow2$) drop into ultraviolet or X-ray range,
not seen in the range of visible light spectrum. By modern techniques in
analyzing laser optical spectrum, these experiments are believed 
realizable. 

From (5.3), the positive ground state energy for $n=1$ and $j=1/2$ is 
$$E_0={m\over{\sqrt{1+(Z\alpha)^2}}},\eqno(5.6)$$
which approaches zero when $Z\rightarrow\infty$. While the
conventional ground state 
energy of a Hydrogen-like ion has a singularity at $Z\alpha\sim1$: 
$$E_{con}=m\sqrt{1-(Z\alpha)^2},\eqno(5.7)$$
which has no real meaningful solution when $Z\alpha>1$.
The percentage difference between (5.6) and (5.7) is 
$$1-{{E_{con}}\over{E_0}}=1-\sqrt{1-(Z\alpha)^4},\eqno(5.8)$$
which is 1\% when $Z\sim50$, 5\% when $Z\sim75$ and 15\% when $Z\sim100$. One
can see that the errors are significant only for high $Z$ 
Hydrogen-like ions which are unstable on the other hand. This causes real
difficulty in doing experiments. An alternative way is to measure the ground
states of high $Z$ atoms instead of ions, by assuming the outer shell
electrons have little screening effects on the inner ground state electrons,
which may be approximately described by single-electron model. So a
possible experiment is to use laser beam to pump out the electrons of
high $Z$ atoms. One may find out the largest ionization energies and compare
them with the theoretical results (5.6) and (5.7). 

In summary, the energy solutions (5.3) differ from conventional results in a subtle way. The differences may not be noticeable if no special attention is paid to it in experiment. In fact, the fine structure splitting is exactly the same as usual. Only the energy order in a certain level $n$ of Hydrogen-like ions is reversed. This is why we emphasize the importance of a thorough experimental investigation of the Zeeman effects of Hydrogen-like ions. Ironically no such experiment can be found by far.    

\section* {6. PARTICLE CONJUGATION} 
After a simple check, we find that charge conjugation defined by
C$\Psi=i\gamma^2\Psi^*$, does not
turn (4.23) into the other one with opposite charge. So we need to try
something else. Now making a complex conjugation $*$ on both sides of (4.23),
we get
$$-{\gamma^\mu}^*P_\mu\Psi^*(x)=(m+eJ^\mu A_\mu)\Psi^*(x).\eqno(6.1)$$
Considering ${\gamma^2}^*=-\gamma^2$ and ${\gamma^i}^*=\gamma^i(i=0,1,3)$,
we may pick a unitary transformation
$${\rm O}_s=i\gamma^0\gamma^1\gamma^3,\eqno(6.2)$$ 
which is a complex conjugation operator in spinor space: 
$${\rm O}_s{\gamma^\mu}^*{\rm O}_s^{-1}=\gamma^\mu.\eqno(6.3)$$
Making transformation $-{\rm O}_s$ on both sides of (6.1), we get
$$\gamma^\mu P_\mu({\rm O}_\omega\Psi(x))=(-m-eJ^\mu A_\mu)({\rm O}_\omega\Psi(x)),\eqno(6.4)$$
where ${\rm O}_\omega$ is a complex conjugation in common space, defined by
$${\rm O}_\omega\Psi(x)=i\gamma^0\gamma^1\gamma^3\Psi^*(x).\eqno(6.5)$$
Comparing (6.4) with (4.23), we find that ${\rm O}_\omega$ changes a particle
with 
charge $e$ and mass $m$ into another one with opposite charge $-e$ and
opposite mass $-m$. So we like to call ${\rm O}_\omega$ ``particle
conjugation''. 

In \S3, we have mentioned that common momentum has two eigenvalues
$\pm m$. So we may just as well write another equation:
$$\gamma^\mu P_\mu\Psi_{-m}(x)=-m\Psi_{-m}(x),\eqno(6.6)$$ 
which gives the same type of solutions with the same energy spectrum
as the Dirac
equation (3.5). Adding an 
interaction term, we get an equation similar to (4.23) but with 
``opposite'' mass: 
$$\gamma^\mu P_\mu\Psi_{-m}(x)=(-m+eJ^\mu A_\mu)\Psi_{-m}(x),\eqno(6.7)$$
which can be transformed under ``particle conjugation'' into  
$$\gamma^\mu P_\mu({\rm O}_\omega\Psi_{-m}(x))=(m-eJ^\mu A_\mu)({\rm O}_\omega\Psi_{-m}(x)).\eqno(6.8)$$
Clearly (6.8) and (4.23) show opposite charge but same mass.

Negative charge has been found in experiment long ago, but no negative
mass. In present-day cosmology, the existence of negative mass is
highly controversial. Somehow it involves the precise definition of mass. It
would be naive to draw any definitive 
conclusion just from this 
single-electron model. Hence we consider this ``particle conjugation''
as a speculative thought, though we do feel equations (4.23), (6.4),
(6.7) and (6.8) are all legitimate forms in certain sense.

Due to our different definitions of unitary time inversion and ``particle
conjugation'', the conventional CP or CPT theorem does not hold in our
theory. It is possible to find more conventional theorems that may not
be derivable in our nonlinear theory. We may either derive similar ones as
replacements, or find completely new theorems.  

\section*{7. PERTURBATION TREATMENT}
As long as the interaction potential $V(x)$ is weak enough in (4.25), we may follow 
Feynman's perturbation treatment (Feynman [23]), to derive the lowest-order
differential cross section: 
$${{d\sigma}\over{d\Omega}}={{|{\bf p_f}|^3}\over{|\bf
p_i|}}{{m^2}\over{E_fE_i}}|\overline
u(p_f,s_f^0)u(p_i,s_i^0)|^2|V(q)|^2,\eqno(7.1)$$ 
where subscripts $i$ and $f$ represent the incoming and outgoing waves
respectively, $\overline u$ and $u$ are the spinors,  $q=p_f-p_i$ is the
four-momentum transfer and 
$$V(q)=\int d^4xV(x)e^{iq\cdot x},\eqno(7.2)$$
is the Fourier amplitude of the interaction potential $V(x)$. 

In the case of Coulomb interaction between the incident electron
beam and nuclein target of nuclear number $Z$: $V(x)=-Z\alpha/|{\bf x}|$, 
energy is conserved: $E_f=E_i$. We have $V(q)=-Z\alpha/{\bf q}^2$
where ${\bf q}^2=({\bf p_f-p_i})^2=-4{\bf p}^2(\sin{\theta\over2})^2$.
For any unpolarized incident electron beam, the cross section is a sum
over final spin states and an average over initial spin states. So we get 
$${{d\overline\sigma}\over{d\Omega}}={{Z^2\alpha^2m^2}\over{{\bf
q}^4}}\Sigma_1.\eqno(7.3)$$
where the spin sum
$$\Sigma_1={1\over2}\sum_{s_f^0,s_i^0}|\overline
u(p_f,s_f^0)u(p_i,s_i^0)|^2={1\over2}(1+{{p_i\cdot p_f}\over{m^2}}),\eqno(7.4)$$
is different from the conventional result (Bjorken and Drell
[7]) since our Coulomb term as in (5.2) does not have a matrix $\beta$. Ignoring a constant factor 1/16, finally we have  
$${{d\overline\sigma}\over{d\Omega}}={{Z^2\alpha^2m^2}\over{{\bf
p}^4(\sin{\theta\over2})^4}}[1+{{{\bf
p}^2}\over{m^2}}(\sin{\theta\over2})^2].\eqno(7.5)$$
In the non-relativistic case $|{\bf p}|\ll m$, it reduces to the Rutherford
formula: 
$${{d\overline\sigma}\over{d\Omega}}={{Z^2\alpha^2m^2}\over{{\bf
p}^4(\sin{\theta\over2})^4}}.\eqno(7.6)$$
And in the relativistic limit $|{\bf p}|\gg m$, it turns out to be 
$${{d\overline\sigma}\over{d\Omega}}={{Z^2\alpha^2}\over{{\bf
p}^2(\sin{\theta\over2})^2}},\eqno(7.7)$$
which differs from the Mott [24] formula by a factor
$[\cot(\theta/2)]^2$. 

Usually the Coulomb interaction is applicable only in low
energy scattering, not in high energy
scattering due to the recoil of nuclear target. In the
electron-proton elastic scattering experiments (McAllister and Hofstadter
[25]), the incident electron 
energy 188 Mev is in the same order as the proton mass 938 Mev. Considering the
recoil effect of hydrogen, we would better utilize a current-potential
interaction with a nonsingular convolution (4.24): 
$$V(x)=eJ_e^\mu(x)*A_\mu^p(x).\eqno(7.8)$$
Its Fourier amplitude becomes 
$$V(q)=e\int d^4x'J_e^\mu(x')e^{iq\cdot x'}\int
d^4xA^p_\mu(x-x')e^{iq\cdot(x-x')}=eJ_e^\mu(q)A_\mu^p(q).\eqno(7.9)$$
From the Maxwell equation (4.18), we may express 
the four-potential driven by proton as follows:   
$$A_\mu^p(x)=G(x)*e_pJ_\mu^p(x)=\int d^4yG(x-y)e_pJ_\mu^p(y),\eqno(7.10)$$
where $G(x)$ is Green's function defined by (4.22). In analogy to (7.9),
we get 
$$A_\mu^p(q)=e_pG(q)J_\mu^p(q),\eqno(7.11)$$
where $G(q)=-(q^2+i\epsilon)^{-1}$. Finally we write down 
$$V(q)=-e^2J_e^\mu(q)G(q)J_\mu^p(q)$$
$$=-\alpha\sqrt{{m^2}\over{E_fE_i}}\sqrt{{M^2}\over{E_f^pE_i^p}}\overline
u(p_f,s_f)\gamma^\mu u(p_i,s_i){{-1}\over{q^2+i\epsilon}}\overline
u(P_f,S_f)\gamma_\mu u(P_i,S_i).\eqno(7.12)$$

In the relativistic elastic scattering as shown in Fig. 1, the four-momentum transfer
squared is 
$q^2=(p'-p)^2=-4E_fE_i(\sin{\theta\over2})^2$, 
and the finite mass recoil factor is 
$E_f/E_i=[1+(2E/M)(\sin{\theta\over2})^2]^{-1}$ where $E=E_i$. For any
unpolarized electron beam, the electron spin is random over
space-time. So when we calculate the interaction
between electron and proton, their spins are chosen randomly. The incoming
electron spin $s_i$ in calculating the Feynman diagrams is not related to the
incident electron spin $s_i^0$ in the past. Same is true for
the outgoing electrons. To calculate the total differential cross section per
solid angle,
we sum over all spin states independently:
$${{d\overline\sigma}\over{d\Omega}}={{\alpha^2m^4E_f}\over{q^4E_i^3}}
\Sigma_1\Sigma_2,\eqno(7.13)$$
where $\Sigma_1$, an extra term that the conventional theory does not
have, is given by (7.4):
$\Sigma_1=(E_fE_i/m^2)(\sin{\theta\over2})^2$ when $E\gg m$, 
and $\Sigma_2$ is the same as in the conventional theory (Bjorken and
Drell [7]): 
$$\Sigma_2={1\over4}\sum_{s_f,s_i}\sum_{S_f,S_i}|\overline
u(p_f,s_f)\gamma^\mu u(p_i,s_i)\overline u(P_f,S_f)\gamma_\mu u(P_i,S_i)|^2$$
$$={{E_fE_i}\over{m^2}}(\cos{\theta\over2})^2[1-{{q^2}\over{2M^2}}(\tan
{\theta\over2})^2].\eqno(7.14)$$ 
Combining all these results we obtain
$${{d\overline\sigma}\over{d\Omega}}={{\alpha^2}\over{E^2}}(\cot{\theta\over2})^2
[{{1-{{q^2}\over{2M^2}}(\tan{\theta\over2})^2}\over{1+{{2E}\over
M}(\sin{\theta\over2})^2}}],\eqno(7.15)$$
which differs from the conventional result by a factor
$[\sin(\theta/2)]^{-2}$. 

In comparison with the experimental data (McAllister and Hofstadter [25])
as shown in Fig. 2,
it is hard to tell which one is better. More advanced research need 
to be made. Generally two form factors need to be considered due to the
finite size and 
anomalous magnetic moment of proton (Cahn and Goldhaber [26]). The
modification due to the internal structure of proton is 
significant in high energy scattering processes: the higher the incident
electron energy, the larger the modification. In the extreme
relativistic limit $|{\bf p}|\gg M$, the result (7.15) is not valid, and
in the  
certain range $|{\bf p}|\sim M$, it is not accurate, but in the low
relativistic limit $m\ll|{\bf p}|\ll M$, it may be good enough. The lowest-order
scattering is quite preliminary. So we would not draw any 
definitive conclusions as yet. 

\section*{8. GAUGE INVARIANCE}
From the correspondence (4.17) and the following relation
$$\overline\Psi(\Phi^{\rm I}-\mbox{\boldmath$\gamma$}\cdot{\bf A^I})\Psi=\overline\Psi
J^\mu A_\mu\Psi=\overline\Psi\gamma^\mu(\overline\Psi\Psi)A_\mu\Psi,\eqno(8.1)$$ 
we can also write (4.1) or (4.23) as follows: 
$$\gamma^\mu(P_\mu-eA_\mu^{\rm T})\Psi=m\Psi,\eqno(8.2)$$
by introducing an interaction four-potential
$$A_\mu^{\rm T}=(\overline\Psi\Psi)A_\mu.\eqno(8.3)$$ 
Here $A_\mu^{\rm T}$ is ``time-covariant'', since $A_\mu$ is ``time-anticovariant'' and $\overline\Psi\Psi$ changes sign under
unitary time inversion. The equation (8.2) implies a formal similarity
to the
Dirac equation (3.5) 
if we define an ``effective'' time-covariant four-momentum 
$$P_\mu^{eff}=P_\mu-eA_\mu^{\rm T},\eqno(8.4)$$
or an ``effective'' time-covariant four-derivative 
$$\partial_\mu^{eff}=\partial_\mu+ieA_\mu^{\rm T}.\eqno(8.5)$$
 
In classical electrodynamics, observable physical quantities can be expressed
in terms of electromagnetic field strengths:
${\bf E}=-\mbox{\boldmath$\nabla$}\Phi-\partial_t{\bf A}$ and ${\bf
B}=\mbox{\boldmath$\nabla$}\times{\bf A}$
which seem invariant under gauge transformation for an arbitrary scalar
function $f(x)$ (Bjorken and Drell [27]; Aithison and Hey [28]):  
$$A_\mu\rightarrow A_\mu+\partial_\mu f.\eqno(8.6)$$ 
However there are several restrictions that need to be taken into account
seriously, if we want to preserve both gauge invariance and Lorentz
invariance of the Maxwell field equation and relativistic fermion
field equation. 
 
First of all, as we have pointed out, Minkowski four-potential $A_\mu$ is
time-anticovariant (3.12) but four-derivative $\partial_\mu$ is apparently
time-covariant under time inversion: 
$$(\partial_t,\partial_{\bf x})\rightarrow(-\partial_t,\partial_{\bf x}).\eqno(8.7)$$
So gauge function $f(x)$ must change sign under time inversion to keep the
transformed four-potential $A_\mu+\partial_\mu f$ time-anticovariant
as same as $A_\mu$. Second,
gauge function $f(x)$ needs to be a proper Lorentz invariant to ensure the
transformed four-potential $A_\mu+\partial_\mu f$ covariant under the
proper Lorentz transformations. 
Third, by making a local phase transformation on fermion field
$$\Psi\rightarrow\Psi\exp{[-i\alpha(x)]},\eqno(8.8)$$ 
and a local gauge transformation (8.6) on Minkowski field simultaneously in
(8.2), we arrive at another constraint 
$$\partial_\mu\alpha-e(\overline\Psi\Psi)\partial_\mu f=0.\eqno(8.9)$$
The general solutions satisfying the above three conditions are 
$$f(x)=\sum C_n(\overline\Psi\Psi)^n, \hskip0.05in (n=\pm1,\pm3,\pm5,\dots),\eqno(8.10a)$$
$$\alpha(x)=e(\sum {n\over{n+1}}C_n(\overline\Psi\Psi)^{n+1}-C_{-1}\log|\overline\Psi\Psi|), \hskip0.05in (n=1,\pm3,\pm5,\dots),\eqno(8.10b)$$
where $C_n$ are arbitrary real constants to cover the whole phase space. Note: $\overline\Psi\Psi\not=0$, otherwise, there is no interaction term in the equation (8.2). 

Furthermore, the Lorentz gauge (4.19) leads to another constraint on the
choice of gauge function $f(x)$:
$$\partial_\mu\partial^\mu f(x)=0.\eqno(8.11)$$ 
Using (4.23) or more generally (4.25) with (4.24), we can prove step
by step that this is true if the gauge function takes the form of (8.10a). 
First we have an expansion 
$$\partial_\mu\partial^\mu(\overline\Psi\Psi)=\overline\Psi(\partial_\mu
\partial^\mu\Psi)+(\partial_\mu\partial^\mu\overline\Psi)\Psi+2\partial_\mu
\overline\Psi\partial^\mu\Psi.\eqno(8.12)$$
From (4.25) it comes out 
$$\gamma^\mu\partial_\mu\Psi=-i(m+V)\Psi.\eqno(8.13)$$
Making a hermitian conjugation on both sides and then multiplying
it by $\gamma^0$ from right, we get by the relations
${\gamma^\mu}^\dagger\gamma^0=\gamma^0\gamma^\mu$ and
$\overline\Psi=\Psi^\dagger\gamma^0$
$$\partial_\mu\overline\Psi\gamma^\mu=i(m+V)\overline\Psi.\eqno(8.14)$$  
From (8.13) it is derived 
$$\partial_\mu\partial^\mu\Psi=\gamma^\mu\partial_\mu\gamma^\nu\partial_\nu
\Psi=-i\gamma^\mu(\partial_\mu V)\Psi-(m+V)^2\Psi.\eqno(8.15)$$
Repeating the procedures in deriving (8.14) leads to
$$\partial_\mu\partial^\mu\overline\Psi=i\overline\Psi\gamma^\mu\partial_\mu 
V-(m+V)^2\overline\Psi.\eqno(8.16)$$ 
Furthermore we have
$$2\partial_\mu\overline\Psi\partial^\mu\Psi=2\partial_\mu\overline\Psi
g^{\mu\nu}\partial_\nu\Psi=\partial_\mu\overline\Psi(\gamma^\mu\gamma^\nu+
\gamma^\nu\gamma^\mu)\partial_\nu\Psi.\eqno(8.17)$$
Combining (8.13) and (8.14), we get
$$\partial_\mu\overline\Psi\gamma^\mu\gamma^\nu\partial_\nu\Psi=(m+V)^2
\overline\Psi\Psi.\eqno(8.18)$$
Then we need to find the second term in (8.17). Now let us make a 
transformation on all fermion fields (note $\gamma^0{\gamma^\mu}^\dagger\gamma^0=\gamma^\mu$) 
$$\Psi'=\gamma^\mu\Psi,\;\overline\Psi'=\overline\Psi\gamma^\mu,\eqno(8.19)$$
then (8.13) becomes 
$$\gamma^\nu\partial_\nu(\gamma^\mu\Psi)=-i(m+V')\gamma^\mu\Psi.\eqno(8.20)$$
Here $V'$ is the interaction after the transformation, which for
example can be expressed by its Fourier amplitude (7.13) in the electron-proton
scattering
$$V'(x)=-\alpha\sqrt{{m^2}\over{E_fE_i}}\sqrt{{M^2}\over{E_f^pE_i^p}}\int{{d^4q}
\over{(2\pi)^4}}{{-1}\over{q^2+i\epsilon}}e^{-iq\cdot x}$$
$$\times\overline u(p_f,s_f)\gamma^\mu\gamma^\rho\gamma^\mu
u(p_i,s_i)\overline u(P_f,S_f)\gamma^\mu\gamma_\rho\gamma^\mu u(P_i,S_i).\eqno(8.21)$$
Given $\gamma^\mu\gamma^\mu=\pm1$, it is clear to see (8.21) does not change when $\gamma^\mu$ and
$\gamma^\rho$ switch twice in both electron and proton (or any
fermion) currents. So the interaction will not change after the
transformation (8.19):  
$$V'(x)=V(x).\eqno(8.22)$$
Combining (8.20), (8.22) and (8.14), we have 
$$\partial_\mu\overline\Psi\gamma^\nu\gamma^\mu\partial_\nu\Psi=-i(m+V)\partial_\mu
\overline\Psi\gamma^\mu\Psi=(m+V)^2\overline\Psi\Psi.\eqno(8.23)$$
With (8.18) and (8.23), then (8.17) becomes 
$$2\partial_\mu\overline\Psi\partial^\mu\Psi=2(m+V)^2\overline\Psi\Psi.\eqno(8.24)$$
Finally inserting (8.15), (8.16) and (8.24) into (8.12) we arrive at 
$$\partial_\mu\partial^\mu(\overline\Psi\Psi)=0,\eqno(8.25)$$
leading to the constraint (8.11) if the gauge function is given by
(8.10a). From the above proof, we can see that the gauge condition is quite stringent.
It appears that the Lorentz gauge is inherent in our theory. Note: the
Coulomb gauge, as a special case of the Lorentz gauge for static
fields, is included.

In summary, with the explicit solutions
(8.10) for $f(x)$ and $\alpha(x)$, the local gauge 
transformations in our nonlinear QED look like: 
$$\Psi\rightarrow\Psi\exp{[-i\alpha(x)}],\eqno(8.26a)$$
$$A_\mu\rightarrow A_\mu+\partial_\mu f(x),\eqno(8.26b)$$
$${\rm or}\; A_\mu^{\rm T}\rightarrow
A_\mu^{\rm T}+{1\over e}\partial_\mu\alpha(x),\eqno(8.26c)$$ 
which ensure both gauge invariance and Lorentz invariance of the
nonlinear equation (8.2) and the Maxwell equation with the Lorentz gauge.  Note: our nonlinear QED involving higher-order fermion fields, is
different from the one involving higher-order electromagnetic fields,
and is also different from the one involving self-interaction
potential in the Dirac EM-equation, mentioned in some literatures (see
reference [29]).

We may also construct a
nonlinear QED Lagrangian density 
$$L(x)=\overline\Psi\gamma^\mu(i\partial_\mu-{1\over2}eA_\mu^{\rm
T})\Psi-m\overline 
\Psi\Psi-{1\over8}F_{\mu\nu}F^{\mu\nu}\overline\Psi\Psi,\eqno(8.27)$$
here $F_{\mu\nu}=\partial_\nu A_\mu-\partial_\mu A_\nu$ is the second-rank antisymmetric electromagnetic field tensor. 
The Hamiltonian density changes sign under unitary time inversion, so
does the 
Lagrangian density. All Euler-Lagrange field equations can be obtained
by Hamilton's principle.  
Taking an infinitesimal arbitrary variation on the electromagnetic field $\delta
A_\mu$, we get
$$\partial^\nu[F_{\mu\nu}(\overline\Psi\Psi)]=e(\overline\Psi\gamma_\mu\Psi)(\overline\Psi\Psi),\eqno(8.28)$$
then keeping the fermion field unchanged
$\partial^\mu(\overline\Psi\Psi)\rightarrow0$, we eliminate 
$F_{\mu\nu}\partial^\nu(\overline\Psi\Psi)$ from (8.28) and cancel $\overline\Psi\Psi$
on both sides to obtain the Maxwell equation   
$\partial^\nu F_{\mu\nu}=eJ_\mu$
that reduces to (4.18) under the Lorentz gauge (4.19). Similarly by taking an
infinitesimal change on the fermion field $\delta\overline\Psi$, we get
$$i\gamma^\mu\partial_\mu\Psi-{1\over2}e\gamma^\mu
A_\mu^{\rm T}\Psi-{1\over2}e(\overline\Psi\gamma^\mu\Psi)A_\mu\Psi-m\Psi-{1\over8}F_{\mu\nu}F^{\mu\nu}\Psi=0,\eqno(8.29)$$
then keeping the electromagnetic field unchanged $\partial_\nu
A_\mu\rightarrow0$, we eliminate the last term in (8.29) 
to obtain the nonlinear equation (8.2) by the relation (8.1). 

Though the gauge transformation (8.26) does not keep the Lagrangian density
(8.27) invariant, it does keep invariant the integral of the Lagrangian
density since the extra term in the integral makes no contribution:
$\int d^4xJ^\mu\partial_\mu\alpha=\int d^4x\partial_\mu(J^\mu\alpha)=0$.
With the definition (8.3), the Maxwell equation (4.18) and the Lorentz gauge
(4.19) or (8.25), we get
$$\partial_\rho\partial^\rho
A_\mu^{\rm T}=(\overline\Psi\Psi)J_\mu^{ext}+2\partial_\rho(\overline\Psi\Psi)\partial^\rho
A_\mu.\eqno(8.30)$$
Suppose in the free field case ($J_\mu^{ext}=0$), the fluctuations of fermion
field and photon field are small, namely 
$$\partial_\rho(\overline\Psi\Psi)=\epsilon_\rho(x)(\overline\Psi\Psi),\eqno(8.31a)$$
$$\partial^\rho A_\mu=\kappa^\rho(x)A_\mu,\eqno(8.31b)$$
where $\epsilon_\rho(x)$ and $\kappa^\rho(x)$ are small fluctuation
functions. Then we end up with 
$$\partial_\rho\partial^\rho A_\mu^{\rm T}+\mu^2(x)A_\mu^{\rm T}=0,\eqno(8.32a)$$
$$\mu^2(x)=-2\epsilon_\rho(x)\kappa^\rho(x).\eqno(8.32b)$$
If $\mu^2(x)>0$, it is a Klein-Gordon equation for vector boson with a
``fluctuation mass'' $\mu(x)$, i.e, a harmonic oscillator with a fluctuation
energy-momentum. If $\mu^2(x)<0$, it shows the decay and oscillation
of the
interaction field. Both cases are complicated: the time-covariant
intermediate vector boson field $A_\mu^{\rm T}$, namely the coupled interaction
field between the Dirac field $\Psi$ and the Minkowski field $A_\mu$, 
fluctuates in space-time.

By (8.27) and (8.1), the interaction Hamiltonian density can be written as 
$$H^{\rm I}(x)={1\over2}e\overline\Psi\gamma^\mu
A_\mu^{\rm T}\Psi={1\over2}e\overline\Psi J^\mu A_\mu\Psi.\eqno(8.33)$$
If the fluctuation is small, we may simply take
zeroth-order approximation by setting $\mu(x)\rightarrow0$. In this limit, we
may treat $A_\mu^{\rm T}$ as a time-covariant massless photon and linearize our
nonlinear QED to fit experimental
data just as accurately as conventional linear QED. Furthermore, we may
investigate the nonlinear aspects of the theory, 
by considering $A_\mu^{\rm T}$ as a massive intermediate boson, or
considering $J^\mu A_\mu$ as a nonsingular 
convolution shown in (4.24) to avoid singularities, and rewrite (8.27) as:
$$L(x)=\overline\Psi(i\gamma^\mu\partial_\mu-{1\over2}eJ^\mu* A_\mu)\Psi-m\overline 
\Psi\Psi-{1\over8}F_{\mu\nu}F^{\mu\nu}\overline\Psi\Psi.\eqno(8.34)$$
With the basic field quantization
techniques, we can then deal 
with various problems in nonlinear QED starting from this Lagrangian density. 

Orthodox QED, on the basis of the Dirac EM-equation, is just a linear
theory. In this linear QED, electron is point-like and the
self-energy of electron blows up when its size reduces to zero.  Also at each
vortex in Feynman diagrams, there is a singular term that causes
ultraviolet 
divergences and requires renormalization. In contrast, a nonlinear equation of the type (4.25) with (4.24) gives
certain ``soliton-like'' solutions,  
not ``point-like'' solutions. So the self-energy of
electron would not go to infinity in our nonlinear QED. It appears at first
sight that this theory is not renormalizable for there is a
fourth-power term 
of fermion field in the Lagrangian. However the interaction potential $eJ^\mu* A_\mu$ in (8.34) is
considered as a nonsingular convolution, each
vortex in Feynman diagrams is smeared, not singular any more as pictured in Fig. 1. Consequently, only
two powers 
are left in the perturbation expansion with the other two being integrated out in
each vortex. Hence this theory is free of ultraviolet divergences and
is also
renormalizable. These are just some general observations. Due to the
scope limit of this paper, I leave these fundamental issues open for future
research. 

So much has been discussed, it is however not my intention to ``solve''
this nonlinear QED in such single paper. Rather my emphasis is on its
``derivation'' by the implementation of  
unitary time inversion. A complete understanding of nonlinear QED entails much
more studies both theoretically and experimentally. It has been
recognized for a long time that nonlinearity 
may result in fundamentally new phenomena (see reference [29] for a
history overview). There is a typical
problem posed for nonlinear theories: since the linear
superposition principle of quantum mechanics is not valid in nonlinear
field equations, it becomes   
quite problematic to make linear perturbation expansions. A great many efforts
have been made in the past several 
decades, to seek non-perturbative techniques in solving nonlinear
quantum field theories. Many interesting and important issues are still far
from being settled.

\section*{9. REMARKS}
The theory established in this paper is strictly limited to the Minkowski flat
space-time, where the Lorentz group is the fundamental symmetry
group. Naturally we would ponder on the possibility of extending our
theory into curved space-time. In
a globally curved space-time, the Lorentz group is only locally
preserved. A conventional treatment on the Dirac fields in such a space-time is
to use a covariant derivative with spin affine connection
in the Dirac equation, as outlined in my dissertation (Jin [30]). As
far as the electromagnetic interaction  
is concerned, a logical step is to include an electro-dynamical interaction
potential rather than external potential in the Dirac equation, similar to
what I have shown here in this paper. After all, a curved space-time
with Lorentzian signature reduces locally to Minkowski space-time as a limit. 
Specifically in a static space-time with a
time-independent metric, the time component
of spin affine connection vanishes, we can 
separate time from space and anticipate a positive-negative symmetric
energy spectrum (Jin [31] [32]). However in a general
space-time with a time-dependent metric, time and space are not separable, time
inversion is not applicable, and it is no longer possible to obtain a
positive-negative symmetric energy spectrum.  

\section* {ACKNOWLEDGMENTS}
The author would like to thank Professor Jonathan Dimock for
helpful discussions. The author is also grateful to many colleagues for
their comments. 

\section*{REFERENCES}
1. E.P. Wigner, {\it Nachr. Akad. Wiss. Gottingen, Math.-Physik}, 546 (1932)\\ 
2. E.P. Wigner, {\it Group Theory} (Academic Press, New York, 1959)\\
3. E.P. Wigner, {\it Symmetries and Reflections} (Indiana University Press,
Bloomington, 1967)\\  
4. A. Einstein, {\it The Principle of Relativity} (Dover, New York, 1952)\\
5. P.A.M. Dirac, {\it Proc. Roy. Soc.} {\bf A117}, 610 (1928); {\bf A118},
351 (1928)\\
6. O. Klein, {\it Z. Phys.} {\bf 53}, 157 (1929)\\  
7. J.D. Bjorken and S.D. Drell, {\it Relativistic Quantum
Mechanics} (McGraw-Hill, New York, 1964)\\
8. P. Zeeman, {\it Phil. Mag.} (5) {\bf 43}, 226 (1897); (5) {\bf 44}, 55, 255
(1897)\\
9. H.A. Lorentz, {\it The Theory of Electrons} (Dover, New York, 1952)\\
10. P.A.M. Dirac, {\it Spinors in Hilbert Space} (Plenum Press, New York, 1974)\\
11. P.A.M. Dirac, {\it The Principles of Quantum Mechanics} (Oxford University Press, London, 1958)\\
12. H. Minkowski, {\it Ann. Phys. Lpz.} {\bf 47}, 927 (1915)\\
13. P.A.M. Dirac, {\it Proc. Roy. Soc.} {\bf A126}, 360 (1930)\\
14. C.D. Anderson, {\it Science} {\bf 76}, 238 (1932);
{\it Phys. Rev.} {\bf 43}, 491 (1933)\\
15. W. Greiner, {\it Relativistic Quantum Mechanics} (Springer-Verlag, Berlin, 1990)\\
16. W. Greiner, B. Muller, and J. Rafelski, {\it Quantum Electrodynamics
of Strong Fields} (Springer-Verlag, Berlin, 1985)\\ 
17. G.N. Lewis and F.H. Spedding, {\it Phys. Rev.} {\bf 43}, 964 (1933)\\
18. F.H. Spedding, C.D. Shane, and N.S. Grace, {\it Phys. Rev.} {\bf 44},
58 (1933)\\
19. F. Paschen, {\it Ann. d. Phys.} {\bf 82}, 692 (1926)\\
20. F. Paschen and E. Back, {\it Ann. d, Phys.} {\bf 39}, 897 (1912);
{\bf 40}, 960 (1913)\\
21. H.E. White, {\it Introduction to Atomic Spectra} (McGraw-Hill, New York,
1934)\\ 
22. H.G. Kuhn, {\it Atomic Spectra} (Academic Press, New
York, 1962)\\
23. R.P. Feynman, {\it Phys. Rev.} {\bf 76}, 749 (1949); {\bf 76}, 769 (1949)\\
24. N.F. Mott, {\it Proc. Roy. Soc.} {\bf A124}, 426 (1929); {\bf A135},
429 (1932)\\
25. R.W. McAllister and R. Hofstadter, {\it Phys. Rev.} {\bf 102}, 851 (1956)\\
26. R.N. Cahn and G. Goldhaber, {\it The Experimental Foundations of Particle 
Physics} (Cambridge University Press, Cambridge, 1989)\\
27. J.D. Bjorken and S.D. Drell, {\it Relativistic Quantum
Fields} (McGraw-Hill, New York, 1965)\\
28. I.J.R. Aitchison and A.J.G. Hey, {\it Gauge Theories in Particle
Physics} (Adam Hilger LTD, Bristol, 1982)\\
29. Asim O. Barut, Alwyn van der Merwe, and Jean-Pierre Vigier, eds.,
{\it Quantum, Space and Time --- The Quest Continues: Studies and
Essays in Honour of Louis de Broglie, Paul Dirac and Eugene Wigner}
(Cambridge University Press, Cambridge, 1984)\\
30. W.M. Jin, {\it Dirac Quantum Fields in Curved Space-Time} (Ph.D.
Dissertation, 1999)\\
31. W.M. Jin, {\it Class. Quantum Grav.} {\bf 15} 3163 (1998) gr-qc/0009009\\
32. W.M. Jin, {\it Class. Quantum Grav.} {\bf 17} 2949 (2000) gr-qc/0009010

\end{document}